\documentclass[11pt]{article}

\usepackage{amsmath}
\usepackage{amssymb}
\usepackage{amsfonts}
\usepackage{amsthm}

\usepackage[utf8x]{inputenc} 

\usepackage{graphicx}
\usepackage[autostyle]{csquotes}
\usepackage[linktocpage]{hyperref}
\usepackage{fullpage}
\usepackage{slashed}
\usepackage{mathtools}
\usepackage{indentfirst}
\usepackage{bm}
\usepackage{mathrsfs} 
\usepackage{authblk}
\usepackage{tikz}
\usepackage{tikz-cd}

\usetikzlibrary{arrows}
\usetikzlibrary{arrows.meta, automata,  positioning, quotes}

\usepackage{subfig}
\usepackage{makecell}
\usepackage{array}
\usepackage{float}

\usepackage{ytableau}

\allowdisplaybreaks

\newtheorem{theorem}{Theorem}[section]

\newtheorem{conjecture}[theorem]{Conjecture}

\newcommand{\intg}{\mathbb{Z}}
\newcommand{\rational}{\mathbb{Q}}
\newcommand{\real}{\mathbb{R}}
\newcommand{\complex}{\mathbb{C}}

\newcommand{\lb}{\left(}
\newcommand{\rb}{\right)}
\newcommand{\lsb}{\left[}
\newcommand{\rsb}{\right]}
\newcommand{\lac}{\left\{}
\newcommand{\rac}{\right\}}

\newcommand{\RN}[1]{%
  \textup{\uppercase\expandafter{\romannumeral#1}}%
}

\newcommand{\swmidarrow}{\tikz \draw[-triangle 90] (0,0) -- +(-.1,-.1);}

\newcommand{\semidarrow}{\tikz \draw[-triangle 90] (0,0) -- +(.1,-.1);}

\newcounter{x}
\newcounter{y}
\newcounter{z}

\newcommand\xaxis{210}
\newcommand\yaxis{-30}
\newcommand\zaxis{90}

\newcommand\topside[3]{
  \fill[fill=gray, draw=black,shift={(\xaxis:#1)},shift={(\yaxis:#2)},
  shift={(\zaxis:#3)}] (0,0) -- (30:1) -- (0,1) --(150:1)--(0,0);
}

\newcommand\leftside[3]{
  \fill[fill=yellow, draw=black,shift={(\xaxis:#1)},shift={(\yaxis:#2)},
  shift={(\zaxis:#3)}] (0,0) -- (0,-1) -- (210:1) --(150:1)--(0,0);
}

\newcommand\rightside[3]{
  \fill[fill=black, draw=black,shift={(\xaxis:#1)},shift={(\yaxis:#2)},
  shift={(\zaxis:#3)}] (0,0) -- (30:1) -- (-30:1) --(0,-1)--(0,0);
}

\newcommand\cube[3]{
  \topside{#1}{#2}{#3} \leftside{#1}{#2}{#3} \rightside{#1}{#2}{#3}
}

\newcommand\planepartition[1]{
 \setcounter{x}{-1}
  \foreach \a in {#1} {
    \addtocounter{x}{1}
    \setcounter{y}{-1}
    \foreach \b in \a {
      \addtocounter{y}{1}
      \setcounter{z}{-1}
      \foreach \c in {1,...,\b} {
        \addtocounter{z}{1}
        \cube{\value{x}}{\value{y}}{\value{z}}
      }
    }
  }
}

\title{Link in $\real \mathbb{P}^3$ and the Topological Vertex}
\author{John Chae}
\affil{ychae@formerstudents.ucdavis.edu}
\date{}  

\begin{document}

\maketitle

\begin{abstract}

We provide the first computations of colored unknots and Hopf link in $\real \mathbb{P}^3$ using both the topological vertex and its refinement. Our approach utilizes the toric Calabi-Yau threefold arising from the geometric transition of the cotangent bundle of $\real \mathbb{P}^3$ under the large $N$ duality. We find that the link invariants are series in the Kahler parameters of the toric Calabi-Yau manifold and the $q$-expansions of the rational functions of the series have positivity property. We conjecture that they are Poincare series of an infinite dimensional link homology theory for links in $\real \mathbb{P}^3$.  We compare our results with that of the $S^3$ and speculate the consequences of the series nature of the invariants.

\end{abstract}

\tableofcontents

\section{Introduction}

One of the primary goals of knot theory is deeper understanding of invariants of links. There has been tremendous advances towards the goal for links in $S^3$. Beginning from link polynomials ranging from a single variable polynomial such as the Alexander polynomial, (colored) Jones polynomial, $sl(N)$ polynomial to a two variable HOMFLY polynomial, homological invariants have entered the scene under the name of categorification program (see \cite{KL2, Web, G, NO, S, A} for surveys). They provide insights into the polynomial invariants and new information of links. For example, the graded Euler characteristic of the knot Floer homology~\cite{OS} is the Alexander polynomial and its generalizations to link and the multivariable Alexander polynomial is given by the link Floer homology theory~\cite{OS2}, whose graded Euler characteristic is the multivariable Alexander-Conway polynomial. Furthermore, the Jones polynomial is shown to be the graded Euler characteristic of the Khovanov homology~\cite{K, K2} whereas the $sl(N)$ polynomial can be categorified as the $sl(N)$ Khovanov-Rozansky homology~\cite{KR} and the a two variable HOMFLY polynomial is the graded Euler characteristic of the HOMFLY homology~\cite{KR2}.\\

The categorification program  not only has elucidated of the polynomial nature of the invariant and integrality of its coefficients, it established the connections to mathematical physics. For example, physics realization of the Khovanov homology and the $sl(N)$ Khovanov-Rozansky homology were constructed, respectively in \cite{W} and in \cite{GSV}. Moreover, the graded Euler characteristic of $sl(N)$ link homology theory for links was first computed via the (refined) topological vertex~\cite{AKMV, IKV} in \cite{GIKV}. Their conjectural results were recently confirmed via the colored $sl(N)$ link homology theory~\cite{JW}. This approach associates a noncompact toric Calabi-Yau threefold (TCY) to a 3-manifold in which links are embedded. This class of TCYs admit a graph description in which its geometry is captured by a planar graph consisting of trivalent vertices. They are building blocks for a graph description. Each edge represents a degeneration cycle of a torus fiber of the manifolds. An appropriate gluing of the vertices yields a particular toric Calabi-Yau threefolds. Examples of TCYs are the resolved conifold (see section 2.1), more generally $O(-a) \oplus O(a-2) \rightarrow P^1$, local $P^1 \times P^1$ and local $P^2$~\footnote{Equivalently, it is $ O(-3) \rightarrow P^2$}.\\

Taking advantage of the graphical characterization of noncompact TCYs, a topological invariant and combinatorial object called topological vertex was introduced in \cite{AKMV} (see also \cite{ORV}).  In a representation basis, the topological vertex assigns 2-dimensional partitions to each edge of a toric graph, which play a role of asymptotic boundary conditions for a 3-dimensional partition:
$$
C_{\lambda \mu \nu} (q) = q^{\frac{\kappa(\mu)}{2}} s_{\nu^t}(q^{-\rho}) \sum_{\eta} s_{\lambda^t / \eta} (q^{-\rho-\nu}) s_{\mu / \eta} (q^{-\rho-\nu^t})
$$
where $s_{\mu / \eta}$ is the skew Schur function and $(\lambda ,\mu ,\nu)$ are the 2d boundary partitions (see Appendix A for their conventions). Using gluing rules of the topological vertex, we can compute a partition function of the (A-model) topological string theory whose target space is a TCY. This approach provides the combinatorial interpretation of the partition function. The topological vertex formulation has a variety of applications in both physics and mathematics. In case of former, it was used to verify the equivalence between the topological string theory and the 5d supersymmetric $U(N)$ gauge theories on the level of their partition functions~\cite{IK,IK2,IK3,HIV}.\\

From the mathematical viewpoint, the topological vertex inspired a new direction in the field of enumerative geometry. For example, the topological vertex was defined rigorously in \cite{LLLZ}. It revealed a rich mathematical structure underlying the topological vertex. It was used to find the generating function of an Gromov-Witten (GW) invariants in all genera, which are rational numbers. Furthermore, an important conjecture of TCYs by \cite{GV} was integrality of the so-called Gopakumar-Vafa (BPS) invariants. The conjecture was initially proven in \cite{P} for a certain class of TCYs and then a proof for all TCYs was given in \cite{KO}. These BPS invariants are related to the GW-invariants recursively; the former can be expressed as a linear combinations of the latter. For open topological string sector, the integrality and finiteness of the open BPS invariants were proven in \cite{Y}.\\

Another application of the topological vertex is in topology, which is the theme of our paper. The topological invariance property of the topological vertex allows to compute invariants of links embedded in a closed orientable 3-manifold that are colored by an \textit{arbitrary} finite dimensional irreducible representations. The first such application was done in \cite{GIKV} for colored links in the $S^3$.\\

Although there has been enormous developments for links in $S^3$, links in $\real P^3$ and more generally, Lens spaces have received gradual investigations from the middle of 1990's~\cite{D,KL,CC,CMM,M2} and later in \cite{MA,GA} and then more recently in \cite{MW,C1,C2}.\\

In this paper, we compute colored unknot and Hopf link invariants in $\real P^3$ using the (refined) topological vertex in the A-model of topological string theory. We find a series nature of the link invariant on the level of the rational function coefficients:
$$
\hat{Z}_{\alpha \gamma^t}(Q_b, Q_f, t,q) = \sum_{r=1 , s=0}^{\infty} c_{r,s}(t,q) Q_{b}^r Q_{f}^s \in \rational(q)[[Q_{b}, Q_{f}]],
$$
where
$$
c_{r,s}(t,q) = \begin{cases}
\rational^{\ast}(t,q),\quad r+s \leq |\alpha| + |\gamma|\quad \text{or} \quad r+s > |\alpha| + |\gamma|\quad \&\quad  s>0 \\
0,\quad Otherwise\\
\end{cases}
$$
where the normalized $\hat{Z}_{\alpha \gamma^t} := Z_{\alpha \gamma^t} / Z_{\emptyset \emptyset}$ and $\alpha, \gamma^t$ are the colors of the link. Furthermore, we observe that $q$ expansions of the coefficient functions of the link invariant yield positive integers in both regular and refined cases.\\
\newline

This paper is organized as follows. We first review the large $N$ duality phenomena in the $S^3$ and $RP^3$ in section 2. In section 3, we compute a colored unknot and the Hopf link invariant using the regular topological vertex. We then apply the refined topological vertex to the same links. We finish by comparing our results with that of the $S^3$ in section 5.

\section{Large $N$ Duality for $\real \mathbb{P}^3$}

\subsection{Geometric Transition}

\noindent \underline{$S^3$ Review} We begin by briefly reviewing the conifold transition for $T^{\ast}S^3$~\cite{OV} (see \cite{M} for details). Consider the $SU(N)$ Chern-Simons gauge theory on $S^3$. It was conjectured that this theory is equivalent to (A-type) open topological string theory with target space $M_6 = T^{\ast}S^3$~\cite{W2}. Mathematically, the latter theory is the (open) Gromov-Witten theory~\cite{FP}. The equivalence states that free energy of the Chern-Simons theory in its perturbative regime ($\lambda = g_s N << 1$)~\footnote{$\lambda$ is the t' Hooft coupling and $g_s$ is string coupling constants.} corresponds to a partition function of the open strings propagating in $T^{\ast}S^3$~\footnote{More precisely, the coefficients $C_{g,h}$ in the free energy expansion is equal to the partition function, where $g$ is the genus and $h$ number of boundary components of a worldsheet.} in the limit of large $N$. On $M_6$ side, $N$ number of D2-branes wrap the zero section $S^3$ of $M_6$. At large $N$, $M_6$ undergoes a conifold transition to $O(-1) \oplus O(-1) \rightarrow CP^1$ called the resolved conifold. In this process, $S^3$ collapses to a point and then it grows to Riemann sphere $S^2$. Furthermore, the Lagrangian D-branes disappear. On the resolved conifold side, we have closed topological strings.\\

In the presence of a colored knot $K$ in $S^3$, there is a corresponding knot conormal bundle $T^{\ast}N_K$ in $M_6$. Topologically, $T^{\ast}N_K \approx S^1 \times \real^2$ and it is a (immersed) Lagrangian submanifold of $M_6$. Before the conifold transition, we need to lift $T^{\ast}N_K$ from the zero section. After the transition, $T^{\ast}N_K$ becomes a new Lagrangian submanifold $L_K$ immersed in the resolved conifold. Its construction was shown in \cite{T}. As an application of this picture, a colored unknot invariant was computed in the resolved conifold side, which was shown to agree with that of the Chern-Simons theory side as $N$ is taken to infinity in \cite{OV}.\\ 

Taking advantage of the resolved conifold, alternative method based on the (refined) topological vertex~\cite{AKMV, IKV} was first used in \cite{GIKV}. In addition to 
computations of colored unknots and Hopf links, an existence of a colored link invariant called (normalized) superpolynomial $\bar{P}_{R_1,\cdots, R_l}(L; a,t,q)$ was conjecture~\cite{GIKV}, where $R_1,\cdots, R_l$ are representation of link $L$. Furthermore, \cite{GIKV} proposed that $\bar{P}_{R_1,\cdots, R_l}(L; a=q^N,t,q)$ is a graded Poincare polynomial of $sl(N)$ link homology. And they were shown to be a polynomial in $a$ variable with coefficients as rational functions in $t,q$~\cite{GIKV}. Recently, these results were confirmed using the colored $sl(N)$ link homology~\cite{JW}. There has been further investigations to torus knots~\cite{JKS} and nontorus links~\cite{AKMMM}.\\


\noindent \underline{$\real P^3$} $SU(N)$ Chern-Simons gauge theory on Lens spaces were shown to be equivalent to a hermitian matrix model via the mirror symmetry~\cite{AKMV2}. After geometric transition of $T^{\ast}\real P^3$, it results in a noncompact toric Calabi-Yau 3-fold, local $\complex P^1 \times \complex P^1$. In case of a knot in $\real P^3$, there has been analysis from the Chern-Simons theory side and the B-model topological string theory~\cite{SS}. Furthermore, a mirror symmetry was investigated and proven in the context of the open/closed Gromov-Witten theory for $L(p,1)$ containing a torus knot from the B-model side in \cite{YZ} .  
\newline


\section{Link Invariant}

In this section, we use the toric graph of local $\complex P^1 \times \complex P^1$ (see Figure 1) and the topological vertex to compute power series invariants of the unknot and Hopf link colored by skew-symmetric representations (see Appendix C for symmetric representations).

\subsection{Setup}

\begin{equation}
Z_{\alpha \gamma^t}(Q_b, Q_f, q) = \sum_{\nu_1 , \nu_2} (-Q_b )^{|\nu_1 |+| \nu_2 |} Z_{\alpha \nu_1 \nu_2}(q,Q_f) f_{\nu_1}(q) f_{\nu_2}(q) Z_{\gamma^t \nu_2 \nu_1} (q,Q_f),
\end{equation}
where
$$
f_{\nu_1} = (-1)^{|\nu_1 |} q^{-\frac{\kappa (\nu_1 )}{2}} \qquad f_{\nu_2} = (-1)^{|\nu_2 |} q^{-\frac{\kappa (\nu_2)}{2}}
$$
$$
Z_{\alpha \nu_1 \nu_2}(q,Q_f) = \sum_{\lambda} (-Q_f)^{|\lambda|} C_{\lambda^t \alpha \nu_1}(q) f_{\lambda}(q) C_{\emptyset \lambda \nu_{2}^{t}}(q)
$$
$$
Z_{\gamma^t \nu_2 \nu_1} (q,Q_f) =  q^{\frac{\kappa (\gamma)}{2}}  \sum_{\beta} (-Q_f)^{|\beta|} C_{\gamma^t \beta \nu_1}(q) f_{\beta}(q) C_{\beta^t \emptyset \nu_{2}^{t}}(q)
$$
$$
C_{\lambda^t \alpha \nu_1}  = q^{\frac{\kappa (\alpha)}{2}} S_{\nu_1^t}(q) S_{\lambda}(q^{-\rho -\nu_1}) S_{\alpha}(q^{-\rho -\nu_1^t}),\qquad C_{\emptyset \lambda \nu_{2}^{t}}  = q^{\frac{\kappa (\lambda)}{2}} S_{\nu_2}(q) S_{\lambda}(q^{-\rho -\nu_2}) 
$$
$$
C_{\gamma^t \beta \nu_1}  = q^{\frac{\kappa (\beta)}{2}} S_{\nu_1}(q) S_{\gamma}(q^{-\rho -\nu_1^t}) S_{\beta}(q^{-\rho -\nu_1}),\qquad C_{\beta^t \emptyset \nu_{2}^{t}}  =  S_{\nu_2^t}(q) S_{\beta}(q^{-\rho -\nu_2}) 
$$
where $q^{-\rho-\chi}= \lac q^{1/2 -\chi_1}, q^{3/2 -\chi_2} , q^{5/2 -\chi_2} , \cdots \rac$. We note that a framing factor $q^{\kappa (\gamma)/2}$ was added to make the expression symmetric in the colors of the link components.
\newline



\begin{figure}
\centering
\begin{tikzpicture}
\draw (2,0) -- (0,0);
\draw (0,0) --  (0,2);
\draw (0,2) --  (2,2);
\draw (2,2) --  (2,0);
\draw (0,0) -- node {\swmidarrow} (-1,-1);
\draw (2,0) --  node {\semidarrow} (3,-1);
\draw (2,2) -- node {\swmidarrow} (3,3);
\draw (0,2) --  (-1,3);
\draw (-1,2) -- (-0,3);

\path[-Stealth]

        (2,0) edge["$\nu_2 $"] (0,0)
				
				(0,0) edge["$\lambda$"] (0,2)
				
				(0,2) edge["$\nu_1 $"] (2,2)
				
				(2,2) edge["$\beta $"] (2,0)
				
				(0,2) edge["$\alpha $"] (-1,3);
\end{tikzpicture}
\caption{A toric graph for a Lagrangian brane in the local $\complex P^1 \times \complex P^1$ geometry corresponding to a colored ($\alpha$) unknot in $\real P^3$. The trivial partitions along the three noncompact edges are suppressed.} 
\end{figure}
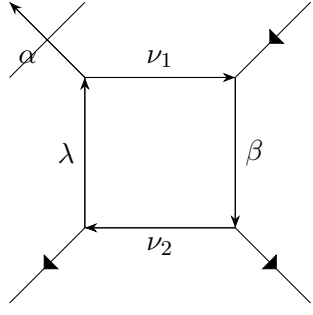

\subsection{Unknot}

For an unknot colored by the fundamental representation in $\real P^3$, we set $\alpha= \square$ and $\gamma=\emptyset$ in (1), we get
\begin{equation}\tag{2}
\begin{split}
Z_{\square \emptyset}(Q,q) & = \sum_{\nu_1 , \nu_2} Q_b^{|\nu_1 |+| \nu_2 |} S_{\square}(q^{-\rho - \nu_{1}^{t}}) q^{-\frac{\kappa (\nu_1)+ \kappa (\nu_2)}{2}} S_{\nu_{1}^{t}} (q^{-\rho}) S_{\nu_2} (q^{-\rho}) S_{\nu_{1}} (q^{-\rho}) S_{\nu_2^{t}} (q^{-\rho}) \\
& \times \sum_{\lambda , \beta} Q_f^{|\lambda |+|\beta |} S_{\lambda} (q^{-\rho - \nu_{1}}) S_{\lambda} (q^{-\rho - \nu_{2}}) S_{\beta} (q^{-\rho - \nu_{1}}) S_{\beta} (q^{-\rho - \nu_{2}})\\
& = \frac{\sqrt{q}}{1-q} + Q_b \,\frac{\sqrt{q}}{1-q} + Q_b Q_f \,\frac{2 \sqrt{q}}{1-q} + \frac{\sqrt{q}}{1-q} \lb 4 Q_{b}^2 Q_f + 3Q_b Q_{f}^2 \rb + O( Q^4),
\end{split}
\end{equation}
where $O( Q^4)$ stands for higher order mixed terms $Q_{b}^m Q_{f}^n, m+n\, \geq 4\, (m,n>0)$.
\newline

\noindent\textbf{Remark}. From the second order in $Q$, no $Q_{b}^2$ and higher power terms appear, which is due to an intricate cancellation.\\
\noindent\textbf{Remark}. We observe that sole $Q_f$ terms do not appear as we expected.\\
\newline
 
\noindent\underline{$q$-Expansion} The series expansion of (2) has manifestly positive integer coefficients as the denominator is a sum of the geometric series.
\newline

\noindent For 2d skew-symmetric representation $\Lambda^2 = \ytableausetup{smalltableaux}\begin{ytableau} ~ \\ ~ \\  \end{ytableau} $.
\begin{equation}\tag{3}
\begin{split}
Z_{\Lambda^2 \emptyset}(Q,q) & = \frac{q}{(1-q) \left( 1-q^2 \right)} + Q_b \, \frac{1}{(1-q)^2} + \frac{1}{(1-q) \left(1-q^2\right)} \lsb Q_{b}^2 \, q + 2Q_b Q_f \, ( q+1) \rsb \\
& + \frac{1}{(1-q)^2} \lsb 3 Q_b Q_{f}^2  + 2 Q_{b}^2 Q_f \, (2+q) \rsb +  O( Q^4)\\
\end{split}
\end{equation}
where $O( Q^4)$ stands for higher order mixed terms $Q_{b}^m Q_{f}^n,\, m+n \geq 4\, (m,n>0)$.
\newline

\noindent\textbf{Remark}. From the third order in $Q$, no $Q_{b}^3$ and higher power terms appear. 
\newline


\noindent\underline{q-Expansion} As in the fundamental representation case, series expansion has positive integer coefficients.
\newline

\noindent For 3d skew-symmetric representation $\Lambda^3 = \ytableausetup{smalltableaux}\begin{ytableau} ~ \\ ~ \\ ~ \\  \end{ytableau} $.
\begin{equation}\tag{4}
\begin{split}
Z_{\Lambda^3 \emptyset}(Q,q) & = \frac{q^{3/2}}{(1-q) \left(1-q^2 \right)\left(1-q^3 \right)} + Q_b \,\frac{1}{q^{1/2}(1-q)^2 (1-q^2)} + \frac{1}{q^{1/2}(1-q)^2 (1-q^2)}\lsb Q_{b}^2  + 2 Q_b Q_f  \rsb \\
& + \frac{1}{q^{1/2}(1-q) \left(1-q^2\right)\left(1-q^3 \right)} \lsb Q_{b}^3 \, q^2 + 2 Q_{b}^2 Q_f \, (3 + 4 q + 4 q^2 +  q^3)  + 3Q_b Q_{f}^2 \, (1 +  q + q^2) \rsb +  O( Q^4)\\
\end{split}
\end{equation}
where $O( Q^4)$ stands for higher order mixed terms $Q_{b}^m Q_{f}^n,\, m+n \geq 4\, (m,n>0)$.  
\newline

\noindent\underline{$q$-Expansion} We observe the positivity property as well. For example,
\begin{equation*}
\begin{split}
Q^0 & : 1+q+2 q^2+3 q^3+4 q^4+5 q^5+7 q^6+8 q^7+10 q^8+12 q^9+14 q^{10}+16 q^{11}+19 q^{12}\\
& +21 q^{13}+24 q^{14}+27 q^{15}+ O\left(q^{16}\right) \in \intg_{+}[[q]]\\
Q_{b}^2 Q_f & : 3+7 q+14 q^2+22 q^3+33 q^4+45 q^5+60 q^6+76 q^7+95 q^8+115 q^9+138 q^{10}+162 q^{11}+189 q^{12}\\
& +217 q^{13}+248q^{14}+280 q^{15} + O\left(q^{16}\right) \in \intg_{+}[[q]]\\
Q_{b} Q_{f}^2 & : 1+2 q+4 q^2+6 q^3+9 q^4+12 q^5+16 q^6+20 q^7+25 q^8+30 q^9+36 q^{10}+42 q^{11}+49 q^{12}\\
& +56 q^{13}+64 q^{14}+72q^{15} + O\left(q^{16}\right) \in \intg_{+}[[q]]
\end{split}
\end{equation*}
\newline


\begin{conjecture} For an unknot $U$ colored by a n-dimensional skew-symmetric or symmetric representation $R$ of $su(2)$ in $\real P^3$, a $q$-series expansion of $Z_{R \emptyset}$
yields positive integer coefficients for all orders in $Q$.
$$
S [Z_{R \emptyset}(U; Q,q)] \in \intg_{+}[[q]],
$$
where $S$ denotes the expansion.
\end{conjecture}

\begin{figure}
\centering
\begin{tikzpicture}
\draw (2,0) -- (0,0);
\draw (0,0) --  (0,2);
\draw (0,2) --  (2,2);
\draw (2,2) --  (2,0);
\draw (0,0) -- node {\swmidarrow} (-1,-1);
\draw (2,0) --  node {\semidarrow} (3,-1);
\draw (2,2) --  (3,3);
\draw (2,2) --  (3,3);
\draw (0,2) --  (-1,3);
\draw (-1,2) -- (-0,3);
\draw (3,2) -- (2,3);
\path[-Stealth]
        (2,0) edge["$\nu_2 $"] (0,0)
				(0,0) edge["$\lambda$"] (0,2)
				(0,2) edge["$\nu_1 $"] (2,2)
				(2,2) edge["$\beta $"] (2,0)
				(0,2) edge["$\alpha $"] (-1,3)
				(3,3) edge["$\gamma^t $"] (2,2);
\end{tikzpicture}
\caption{A toric graph for a pair of Lagrangian branes in the local $\complex P^1 \times \complex P^1$ geometry corresponding to a colored ($\alpha ,\gamma $) Hopf link in $\real P^3$. The trivial partitions along the two noncompact edges are suppressed.} 
\end{figure}
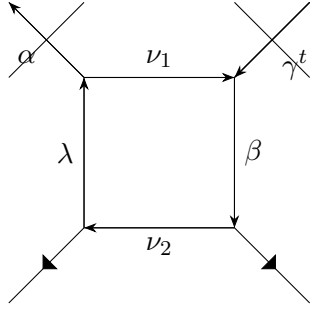

\subsection{Hopf Link}

For an Hopf link colored by the fundamental representations in $\real P^3$, we set $\alpha= \gamma=\square$ in (1).

\begin{equation}\tag{5}\label{}
\begin{split}
Z_{\square \square}(Q,q) & =  \sum_{\nu_1 , \nu_2} Q_b^{|\nu_1 |+| \nu_2 |} ( S_{\square}(q^{-\rho - \nu_{1}^{t}}) )^2 q^{-\frac{\kappa (\nu_1)+ \kappa (\nu_2)}{2}} S_{\nu_{1}^{t}} (q^{-\rho}) S_{\nu_2} (q^{-\rho}) S_{\nu_{1}} (q^{-\rho}) S_{\nu_2^{t}} (q^{-\rho}) \\
& \times \sum_{\lambda , \beta} Q_f^{|\lambda |+|\beta |} S_{\lambda} (q^{-\rho - \nu_{1}}) S_{\lambda} (q^{-\rho - \nu_{2}}) S_{\beta} (q^{-\rho - \nu_{1}}) S_{\beta} (q^{-\rho - \nu_{2}})\\
& = \frac{q}{(1-q)^2} + Q_b \, \frac{1+q^2}{(1-q)^2} +  \frac{1}{(1-q)^2} \lsb Q_{b}^2 \, q  + 2Q_b Q_f \,\left(1+ q^2\right) \rsb + \frac{1}{(1-q)^2} \lsb 4 Q_{b}^2 Q_f \, \lb 1+ q \notag\right. \notag\right.\\
& \notag\left. \notag\left.   + q^2 \rb + 3 Q_{b} Q_{f}^2 \,\lb 1 + q^2 \rb \rsb + O( Q^4)\\
\end{split}
\end{equation}
where $O( Q^4)$ stands for higher order mixed terms $Q_{b}^m Q_{f}^n,\, m+n \geq 4\, (m,n>0)$.
\newline

\noindent\textbf{Remark}. From the third order in $Q$, no $Q_{b}^3$ and higher power terms appear.\\
\newline
\begin{equation}\tag{6}
\begin{split}
Z_{\square \Lambda^2 }(Q,q) & = \frac{q^{3/2}}{(1-q)^2 (1-q^2)} + Q_b \,\frac{1 + q^2 + q^3}{q^{1/2}(1-q)^2 (1-q^2)} +  \frac{1 + q^2 + q^3}{q^{1/2}(1-q)^2 (1-q^2)} \lsb Q_{b}^2   + 2Q_b Q_f  \rsb  \\
&  + \frac{1}{q^{1/2}(1-q)^2 (1-q^2)} \lsb Q_{b}^3 \, q^2 + 2 Q_{b}^2 Q_f \, \lb 3 + 2 q+6 q^2 + 3 q^3 + q^4 \rb + 3 Q_{b} Q_{f}^2 \,(1 +  q^2 +  q^3) \rsb \\
& + O( Q^4)\\
\end{split}
\end{equation}
where $O( Q^4)$ stands for higher order mixed terms $Q_{b}^m Q_{f}^n,\, m+n \geq 4\, (m,n>0)$.
\newline

\noindent\underline{$q$-Expansion} It is clear that the expansions of (5) and (6) yield positive integer coefficients.
\newline
%

\section{Refined Link Invariant}

We use the refined topological vertex to compute normalized invariants of unknots and Hopf links colored by skew-symmetric representations. We find that $q$-series expansions of the normalized invariants yield coefficients in $t$-polynomials with positive integers.\\

Motivated by \cite{AKMV}, one parameter refinement of the topological vertex was introduced in \cite{IKV}:
\begin{equation}\tag{7}
C_{\lambda \mu \nu} (t,q) = \lb \frac{q}{t} \rb^{\frac{||\mu||^2 + ||\nu||^2}{2}} t^{\frac{\kappa (\mu)}{2}} P_{\nu^t}(t^{-\rho};q,t) \sum_{\eta} \lb \frac{q}{t} \rb^{\frac{|\eta|+|\lambda|-|\mu|}{2}} s_{\lambda^t /\eta}(t^{-\rho}q^{-\nu}) s_{\mu /\eta}(q^{-\rho}t^{-\nu^t})
\end{equation}
A topological significance of the refined version is that it connects the topological string theory to link homology theories. Specifically, invariants of colored unknots and Hopf links in $S^3$ were computed based on the toric graph associated with $S^3$ in \cite{GIKV}. Furthermore, the invariants were examples of a  conjectured a superpolynomial, which reduces to a graded Poincare polynomial of $sl(N)$ link homology theory for links in $S^3$~\cite{GIKV}.


\subsection{Setup}

Applying (7) to Figure 2, we obtain the following expression for the refined partition function. 

\begin{equation}\tag{8}
Z_{\alpha,\gamma^t} (Q_b, Q_f, t,q)  = \sum_{\nu_1 , \nu_2} (-Q_b)^{|\nu_1 | + |\nu_2 |} Z_{\alpha \nu_1 \nu_2 }(Q_f,t,q) f_{\nu_1}(t,q) f_{\nu_2}(q,t)  Z_{\gamma^t \nu_1 \nu_2 }(Q_f,q,t)
\end{equation}
\begin{equation*}
\begin{split}
Z_{\alpha \nu_1 \nu_2 } & = \sum_{\lambda} (-Q_f)^{\lambda} C_{\lambda^t \alpha \nu_1}(t,q) f_{\lambda}(t,q) C_{\phi \lambda \nu_2}(t,q)\\
Z_{\gamma^t \nu_1 \nu_2 } & = \sum_{\beta} (-Q_f)^{\beta} C_{\gamma^t \beta \nu_{1}^t}(q,t) f_{\beta}(q,t) C_{\beta^t \phi  \nu_2}(q,t)
\end{split}
\end{equation*}

\begin{equation*}
\begin{split}
C_{\lambda^t \alpha \nu_1}(t,q) & = \lb \frac{q}{t} \rb^{\frac{||\alpha||^2 + ||\nu_1||^2 }{2}} t^{\frac{\kappa (\alpha) + || \nu_1 ||^2}{2}} \tilde{Z}_{\nu_1}(t,q) \lb \frac{q}{t} \rb^{\frac{|\lambda | - |\alpha | }{2}} S_{\lambda}(t^{-\rho} q^{-\nu_1}) S_{\alpha}(q^{-\rho} t^{-\nu_1^t})\\
C_{\phi \lambda \nu_2}(t,q) & = \lb \frac{q}{t} \rb^{\frac{||\lambda||^2 + ||\nu_2^t||^2 }{2}} t^{\frac{\kappa (\lambda) + || \nu_2^t ||^2}{2}} \tilde{Z}_{\nu_2^t}(t,q) \lb \frac{t}{q} \rb^{\frac{|\lambda | }{2}} S_{\lambda}(q^{-\rho} t^{-\nu_2})  \\
C_{\gamma^t \beta \nu_{1}^t}(q,t) & = \lb \frac{t}{q} \rb^{\frac{||\beta||^2 + ||\nu_1^t||^2 }{2}} q^{\frac{\kappa (\beta)}{2}} P_{\nu_1}(q^{-\rho} ;t,q)\lb \frac{t}{q} \rb^{\frac{|\gamma | - |\beta| }{2}} S_{\gamma}(q^{-\rho} t^{-\nu_1^t}) S_{\beta}(t^{-\rho} q^{-\nu_1}) \\
C_{\beta^t \phi  \nu_2}(q,t) & =  \lb \frac{t}{q} \rb^{\frac{||\nu_2||^2 }{2}}  P_{\nu_2^t}(q^{-\rho};t,q )\lb \frac{t}{q} \rb^{\frac{|\beta | }{2}} S_{\beta}(q^{-\rho} t^{-\nu_2})\\
f_{\nu_1}(t,q) f_{\nu_2}(q,t) & = (-1)^{|\nu_1 |} \lb \frac{t}{q} \rb^{\frac{||\nu_{1}^t ||^2 - |\nu_1 |}{2}} q^{-\frac{\kappa (\nu_1)}{2}} (-1)^{|\nu_2 |} \lb \frac{q}{t} \rb^{\frac{||\nu_{2}^t ||^2 - |\nu_2 |}{2}} t^{-\frac{\kappa (\nu_2)}{2}}
\end{split}
\end{equation*}

\subsection{Unknot}

For an unknot colored by the fundamental representation embedded in $\real P^3$, we set $\alpha= \square$ and $\gamma=\emptyset$ in (8).
\begin{equation}\tag{9}
\begin{split}
Z_{\square \emptyset}(\vec{Q},t,q) & = \sum_{\nu_1 , \nu_2} Q_b^{|\nu_1 |+| \nu_2 |} S_{\square}(q^{-\rho} t^{ - \nu_{1}^{t}}) \tilde{Z}_{\nu_1}(t,q) \tilde{Z}_{\nu_{1}^{t}}(q,t) \tilde{Z}_{\nu_2}(q,t) \tilde{Z}_{\nu_{2}^{t}}(t,q) \lb \frac{q}{t} \rb^{\frac{|\nu_1 |-| \nu_2 |}{2}} q^{\frac{||\nu_1||^2 - ||\nu_{1}^{t}||^2 }{2}} t^{||\nu_{1}^{t}||^2} \\
& \times   t^{\frac{||\nu_2||^2 - ||\nu_{2}^{t}||^2 }{2}} q^{||\nu_{2}^{t}||^2} q^{\frac{-\kappa (\nu_1 )}{2}} t^{\frac{-\kappa (\nu_1 )}{2}} \sum_{\lambda , \beta} Q_f^{|\lambda |+|\beta |} \lb \frac{q}{t} \rb^{\frac{|\lambda |+ || \lambda ||^2 - || \lambda^{t} ||^2}{2}} \lb \frac{t}{q} \rb^{\frac{|\beta |+ || \beta ||^2 - ||\beta^{t} ||^2}{2}} \\
& \times \lb \frac{t}{q} \rb^{\frac{\kappa (\lambda) - \kappa (\beta)}{2}} S_{\lambda} (t^{-\rho} q^{- \nu_{1}}) S_{\lambda} (q^{-\rho} t^{- \nu_{2}}) S_{\beta} (t^{-\rho} q^{- \nu_{1}}) S_{\beta} (q^{-\rho} t^{- \nu_{2}})\\
& = \frac{\sqrt{q}}{1-q} + Q_b \,\frac{q}{\sqrt{t}(1-q)} + Q_b Q_f \,\frac{q+t}{\sqrt{t}(1-q)} + \frac{1}{tq(1-q)} \lsb Q_{b}^2 Q_f \, \sqrt{q}\lb q^2 + 2 q t + t^2 \rb \notag\right.\\
& \notag\left. + Q_b Q_{f}^2 \,\sqrt{t} \lb q^2 + q t + t^2 \rb   \rsb + O( Q^4),
\end{split}
\end{equation}
where $O( Q^4)$ stands for higher order mixed terms $Q_{b}^m Q_{f}^n, m+n\, \geq 4\, (m,n>0)$.  As a consistency check, upon setting $t=q$, it reduces to the regular result (2). We notice that $t$ appears as monomials in the denominators, which is not the case in the absence of a knot (see Appendix C). Furthermore, the exchange symmetry between $t$ and $q$ is no longer present in (9). 
\newline

\noindent\underline{$q$-Expansion} For refined case, coefficients of $q$ series expansion are polynomial in $t$ up to overall factors of $t^{1/2}$ and/or $q^{1/2}$. For example,

\begin{equation*}
\begin{split}
Q_b Q_f & : t+(1+t) q +(1+t)q^2 + (1+t)q^3 +(1+t)q^4 +(1+t)q^5 + (1+t)q^6 +(1+t)q^7 \\
& +(1+t)q^8 +(1+t) q^9 + (1+t)q^{10}+(1+t)q^{11}+(1+t)q^{12} +(1+t)q^{13}  + O(q^{14}) \in \intg_{+}[t][[q]]\\
Q_{b}^2 Q_f & : t^2+\left(2 t+t^2\right) q+\left(1+2 t+t^2\right) q^2+\left(1+2 t+t^2\right) q^3 +\left(1+2 t+t^2\right) q^4\\
& +\left(1+2t+t^2\right) q^5+\left(1+2 t+t^2\right) q^6+\left(1+2 t+t^2\right) q^7 +\left(1+2 t+t^2\right) q^8\\
& +\left(1+2t+t^2\right) q^9 +\left(1+2 t+t^2\right) q^{10}+\left(1+2 t+t^2\right) q^{11} +\left(1+2 t+t^2\right)q^{12}\\
& +\left(1+2 t+t^2\right) q^{13}+\left(1+2 t+t^2\right) q^{14}+\left(1+2 t+t^2\right) q^{15} +O\left(q^{16}\right) \in \intg_{+}[t][[q]]\\
Q_{b} Q_{f}^2 & : t^2+\left( t+t^2\right) q+\left(1+ t+t^2\right) q^2+\left(1+ t+t^2\right) q^3 +\left(1+ t+t^2\right) q^4\\
& +\left(1+t+t^2\right) q^5+\left(1+t+t^2\right) q^6+\left(1+t+t^2\right) q^7 +\left(1+t+t^2\right) q^8\\
& +\left(1+t+t^2\right) q^9 +\left(1+ t+t^2\right) q^{10}+\left(1+ t+t^2\right) q^{11} +\left(1+ t+t^2\right)q^{12}\\
& +\left(1+t+t^2\right) q^{13}+\left(1+t+t^2\right) q^{14}+\left(1+ t+t^2\right) q^{15} +O\left(q^{16}\right) \in \intg_{+}[t][[q]]
\end{split}
\end{equation*}
We observe that all the integers in the coefficient functions are positive integers. We predict that this expansion corresponds to the Poincare series of a $sl(N)$ link homology groups for links in $\real P^3$.
\newline

\begin{equation}\tag{10}
\begin{split}
Z_{\Lambda^2 \emptyset}(\vec{Q},t,q) & = \frac{q}{(1-q) \left(1-q^2\right)} + Q_b \, \lb \frac{q}{t} \rb^{3/2} \frac{(1-q+t)}{(1-q)^2} + \frac{1}{t(1-q) \left(1-q^2\right)}\lsb Q_{b}^2 \, q^2 + Q_b Q_f \, \sqrt{ \frac{q}{t} } (q - q^3 \notag\right.\\
& \notag\left.   + t + q t + t^2 + q t^2) \rsb + \frac{1}{(1-q)^2} \lsb Q_b Q_{f}^2 \, \frac{q^2-q^3+q t+t^2+t^3}{t\sqrt{qt} }  + Q_{b}^2 Q_f \,\frac{1}{t^2} \lb q^2-q^3+2 q t \notag\right. \notag\right.\\
& \notag\left. \notag\left. +t^2+2 q t^2+t^3\rb \rsb +  O( Q^4)\\
\end{split}
\end{equation}
where $O( Q^4)$ stands for higher order mixed terms $Q_{b}^m Q_{f}^n,\, m+n \geq 4\, (m,n>0)$. When $t=q$ in (10), we recover (3). We observe that $t$ appears as monomials in the denominators in this example as well.
\newline

\noindent\underline{$q$-Expansion} Up to overall factors of $t^{1/2}$ and/or $q^{1/2}$, we have

\begin{equation*}
\begin{split}
Q_b & : 1+t+ (1+2 t) q + (1+3 t) q^2 +  (1+4 t)q^3 + (1+5 t)q^4 + (1+6 t)q^5 + (1+7 t)q^6 + (1+8 t)q^7 \\
& + (1+9 t)q^8 + (1+10 t)q^9 + (1+11 t)q^{10} + (1+12 t)q^{11} + (1+13 t)q^{12} + (1+14 t)q^{13} \\
& + (1+15 t)q^{14} +(1+16 t)q^{15} + O(q^{16}) \in \intg_{+}[t][[q]]\\
Q_b Q_f & : t (1 + t)  +  (1 + 2 t + 2 t^2) q +  (1 + 3 t + 3 t^2) q^2 +  (1 + 2 t)^2 q^3 +  (1 + 5 t + 5 t^2)q^4\\
& +  (1 + 6 t + 6 t^2)q^5  +  (1 + 7 t + 7 t^2)q^6 +  (1 + 8 t + 8 t^2) q^7 +  (1 + 9 t + 9 t^2)q^8 \\
& +    (1 + 10 t + 10 t^2)q^9 +  (1 + 11 t + 11 t^2)q^{10} +  (1 + 12 t + 12 t^2)q^{11}  +  (1 + 13 t + 13 t^2)q^{12}\\
& +  (1 + 14 t + 14 t^2)q^{13}  + (1 + 15 t + 15 t^2)q^{14}  +   (1 + 16 t + 16 t^2) q^{15} + O(q^{16}) \in \intg_{+}[t][[q]] \\
\end{split}
\end{equation*}
Higher order $Q$-terms are recorded in the appendix. We suppressed $q$-expansions of terms having manifestly positive integer coefficients.
\newline

\begin{equation}\tag{11}
\begin{split}
Z_{\Lambda^3 \emptyset}(\vec{Q},t,q) & = \frac{q^{3/2}}{(1-q) \left(1-q^2\right)\left(1-q^3 \right)} + Q_b \frac{q^2 (1 - q - q^2 + q^3 + t - q^2 t + t^2)}{t^{5/2}(1-q)^2 (1-q^2)} \\
& + \frac{1}{t^3 (1-q)^2 (1-q^2)} \lsb Q_{b}^2 \, q^{5/2}(1 - q - q^2 + q^3 + t - q^2 t + t^2) + Q_b Q_f \, q t^{1/2}(q - q^2 - q^3 \notag\right.\\
& \notag\left.  + q^4 + t - q^2 t + t^2  + q t^2 - q^2 t^2 + t^3)  \rsb\\
&  + \frac{1}{t^3 (1-q) \left(1-q^2\right)\left(1-q^3 \right)} \lsb Q_{b}^3 \, q^3 t^{3/2} +  Q_{b}^2 Q_f \, q^{1/2} (2 q^2 - 2 q^4 - 2 q^5 + 2 q^7 \notag\right.\\
& \notag\left. + (3 q + 3 q^2 - 3 q^4 - 3 q^5) t + (1 + 4 q + 5 q^2 + 2 q^3 -  q^4 - 2 q^5) t^2 + (1 + 4 q + 4 q^2 + 3 q^3) t^3  \notag\right.\\
& \notag\left. + (1 + q + q^2) t^4) + Q_b Q_{f}^2 \, t^{1/2} (q^2 - q^4 - q^5 + q^7 + (q + q^2 - q^4 - q^5) t + (1 + q + q^2) t^2 \notag\right.\\
& \notag\left. + (1 + 2 q + q^2 - q^4) t^3 + (1 + q + q^2) t^4) \rsb +  O( Q^4)\\
\end{split}
\end{equation}
where $O( Q^4)$ stands for higher order mixed terms $Q_{b}^m Q_{f}^n,\, m+n \geq 4\, (m,n>0)$. When $t=q$, (11) reduces to (4). We again notice that $t$ appears as monomials in the denominators.
\newline

\noindent\underline{$q$-Expansion} Up to overall factors of $t^{1/2}$ and/or $q^{1/2}$, we get

\begin{equation*}
\begin{split}
Q_b & : 1 + t + t^2 +  (1 + 2 t + 2 t^2)q +  (1 + 3 t + 4 t^2)q^2 +  (1 + 4 t + 6 t^2)q^3 +  (1 + 5 t + 9 t^2)q^4\\
& + (1 + 6 t + 12 t^2) q^5  +(1 + 7 t + 16 t^2)  q^6 +  (1 + 8 t + 20 t^2)q^7 + (1 + 9 t + 25 t^2)q^8   \\
& +  (1 + 10 t + 30 t^2)q^9 + (1 + 11 t + 36 t^2)q^{10}   + O(q^{11}) \in \intg_{+}[t][[q]]\\
Q_{b}^2 & : 1 + t + t^2 + (1 + 2 t + 2 t^2)q  +(1 + 3 t + 4 t^2) q^2  +  (1 + 4 t + 6 t^2)q^3  + (1 + 5 t + 9 t^2)q^4 \\
&  + (1 + 6 t + 12 t^2) q^5  + (1 + 7 t + 16 t^2)q^6  +  (1 + 8 t + 20 t^2)q^7  + (1 + 9 t + 25 t^2)q^8  \\
&  + (1 + 10 t + 30 t^2) q^9  +  (1 + 11 t + 36 t^2)q^{10} + O(q^{11}) \in \intg_{+}[t][[q]]\\
Q_b Q_f & : t (1 + t + t^2) + (1 + 2 t + 3 t^2 + 2 t^3) q + (1 + 3 t + 5 t^2 + 4 t^3)q^2  + (1 + 4 t + 8 t^2 + 6 t^3)q^3 \\
&  +   (1 + 5 t + 11 t^2 + 9 t^3)q^4 +(1 + 6 t + 15 t^2 + 12 t^3) q^5  +  (1 + 7 t + 19 t^2 + 16 t^3) q^6 \\
& + (1 + 8 t + 24 t^2 + 20 t^3) q^7 +(1 + 9 t + 29 t^2 + 25 t^3) q^8  +(1 + 10 t + 35 t^2 + 30 t^3) q^9  \\
& + (1 + 11 t + 41 t^2 + 36 t^3)q^{10}   + O(q^{11}) \in \intg_{+}[t][[q]]\\
\end{split}
\end{equation*}
Higher order $Q$-terms are listed in the appendix.\\

To the best of author's knowledge, there is no knot theory result, which we can compare the above results to. However, having observed positivity of integrality property of $q$-expansions in the above examples and motivated by the verification of the results in the case of $S^3$ in \cite{GIKV} by \cite{JW}, we state the following conjecture.

\begin{conjecture} For an unknot $U$ colored by a n-dimensional skew-symmetric or symmetric representation $R$ of $su(2)$ in $\real P^3$, when specializing $Q_b$ and $Q_f$ to products of monomial factors in $q$ and $t$, a $q$-series expansion of $Z_{R \emptyset}(U; Q_b, Q_f, t,q)$ is a graded Poincare series of a colored $sl(N)$ link homology theory for links in $\real P^3$ up to an overall factor.  
\end{conjecture}

\noindent\textbf{Remark.} We speculate that powers of the monomial factors depend on $N$.\\
\noindent\textbf{Remark 2.} The proportional factor would depend on the colors of the unknot and framing factors.

\subsection{Hopf Link}

\begin{equation}\tag{12}
\begin{split}
Z_{\square \square}(\vec{Q},t,q) & = \sum_{\nu_1 , \nu_2} Q_b^{|\nu_1 |+| \nu_2 |} \lb S_{\square}(q^{-\rho} t^{ - \nu_{1}^{t}}) \rb^2 \tilde{Z}_{\nu_1}(t,q) \tilde{Z}_{\nu_{1}^{t}}(q,t) \tilde{Z}_{\nu_2}(q,t) \tilde{Z}_{\nu_{2}^{t}}(t,q) \lb \frac{q}{t} \rb^{\frac{|\nu_1 |-| \nu_2 |}{2}} q^{\frac{||\nu_1||^2 - ||\nu_{1}^{t}||^2 }{2}} t^{||\nu_{1}^{t}||^2} \\
& \times   t^{\frac{||\nu_2||^2 - ||\nu_{2}^{t}||^2 }{2}} q^{||\nu_{2}^{t}||^2} q^{\frac{-\kappa (\nu_1 )}{2}} t^{\frac{-\kappa (\nu_1 )}{2}} \sum_{\lambda , \beta} Q_f^{|\lambda |+|\beta |} \lb \frac{q}{t} \rb^{\frac{|\lambda |+ || \lambda ||^2 - || \lambda^{t} ||^2}{2}} \lb \frac{t}{q} \rb^{\frac{|\beta |+ || \beta ||^2 - ||\beta^{t} ||^2}{2}} \\
& \times \lb \frac{t}{q} \rb^{\frac{\kappa (\lambda) - \kappa (\beta)}{2}} S_{\lambda} (t^{-\rho} q^{- \nu_{1}}) S_{\lambda} (q^{-\rho} t^{- \nu_{2}}) S_{\beta} (t^{-\rho} q^{- \nu_{1}}) S_{\beta} (q^{-\rho} t^{- \nu_{2}})\\
& = \frac{q}{(1-q)^2} + Q_b \frac{q^{3/2} }{t^{3/2}}\frac{ 1 - q + t + q t}{ (1 - q)^2 } + \frac{1}{t (1 - q)^2} \lsb Q_{b}^2 q^2  + Q_{b}Q_f \sqrt{\frac{q}{t}} \lb q - q^2 + t + q^2 t \notag\right. \notag\right.\\
& \notag\left.\notag\left.  + t^2 + q t^2 \rb  \rsb + \frac{1}{ t^2 (1 - q)^2 } \lsb Q_b Q_{f}^2 \sqrt{\frac{t}{q}} \lb q^2 - q^3 + q t + q^3 t + t^2 + q^2 t^2 + t^3 + q t^3 \rb \notag\right. \\
&  \notag\left. + Q_{b}^2 Q_{f} \lb q^2 - q^3 + 2 q t + q^2 t + q^3 t + t^2 + 3 q t^2 + 2 q^2 t^2 + t^3 +  q t^3 \rb  \rsb + O (Q^4) \\
\end{split}
\end{equation}
where $O( Q^4)$ stands for higher order mixed terms $Q_{b}^m Q_{f}^n,\, m+n \geq 4\, (m,n>0)$. As a consistency check, setting $t=q$ reduces to (5). We observe that $t$ appears as monomials in the denominators.
\newline

\noindent\underline{$q$-Expansion} Up to overall factors of $t^{1/2}$ and/or $q^{1/2}$, we get

\begin{equation*}
\begin{split}
Q_b & : 1 + t + (1 + 3 t)q  +  (1 + 5 t)q^2 +  (1 + 7 t)q^3 + (1 + 9 t)q^4  +  (1 + 11 t)q^5  +(1 + 13 t) q^6  \\
& + (1 + 15 t)q^7  + (1 + 17 t)q^8  +  (1 + 19 t)q^9  + (1 + 21 t)q^{10}  +  (1 + 23 t)q^{11} +  (1 + 25 t)q^{12} \\
&  +  (1 + 27 t)q^{13} + (1 + 29 t)q^{14}  + (1 + 31 t)q^{15}   + O(q^{16}) \in \intg_{+}[t][[q]]\\
Q_b Q_f & :t + t^2 + (1 + 2 t + 3 t^2) q + (1 + 4 t + 5 t^2)q^2  + (1 + 6 t + 7 t^2) q^3  + (1 + 8 t + 9 t^2) q^4 \\
& + (1 + 10 t + 11 t^2)q^5  + (1 + 12 t + 13 t^2)q^6  +  (1 + 14 t + 15 t^2)q^7  + (1 + 16 t + 17 t^2) q^8 \\
& + (1 + 18 t + 19 t^2)q^9 + (1 + 20 t + 21 t^2)q^{10}  + (1 + 22 t + 23 t^2) q^{11}  +  (1 + 24 t + 25 t^2) q^{12}\\
& + (1 + 26 t + 27 t^2)q^{13} +(1 + 28 t + 29 t^2)  q^{14} +  (1 + 30 t + 31 t^2)q^{15}  + O(q^{16}) \in \intg_{+}[t][[q]]\\
\end{split}
\end{equation*}
Higher order $Q$-terms are recorded in the appendix.
\newline

\begin{equation}
\begin{split}
Z_{\square \Lambda^2 }(\vec{Q},t,q) & = \frac{q^{3/2}}{(1-q)^2 (1-q^2)}  + Q_b \frac{q^{2}(1 - q - q^2 + q^3 + t + q t - q^2 t - q^3 t + t^2 + q t^2 + q^2 t^2)}{t^{5/2} (1-q)^2 (1-q^2)}\\
& + \frac{1}{t^3 (1-q)^2 (1-q^2)} \lsb Q_{b}^2 q^2 \lb 1 - q - q^2 + q^3 + t + q t - q^2 t - q^3 t + t^2 + q t^2 + q^2 t^2 \rb \right. \\
&  \left. + Q_b Q_f q \lb q - q^2 - q^3 + q^4 + t - q^4 t + t^2 + 2 q t^2 + t^3 + q t^3 +  q^2 t^3 \rb \rsb + \frac{1}{t^{3} (1-q)^2 (1-q^2)}\\
& \times \lsb Q_{b}^3 q^3 t^{3/2} + Q_{b}^2 Q_f \sqrt{q} \lb 2 q^2 - 2 q^3 - 2 q^4 + 2 q^5 + 3 q t + q^2 t - q^3 t - q^4 t - 2 q^5 t + t^2 + 4 q t^2 \right.\right.\\
& \left. \left.  + 4 q^2 t^2 + t^3 + 4 q t^3 + 4 q^2 t^3 + 3 q^3 t^3 + t^4 + q t^4 + q^2 t^4 \rb  + Q_b Q_{f}^2 \sqrt{t} \lb q^2 - q^3 - q^4 + q^5 \right.\right.\\
& \left. \left. + q t - q^5 t + t^2 + q^2 t^2 + q^3 t^2 + t^3 + 2 q t^3 + t^4 + q t^4 + q^2 t^4 \rb \rsb  + O( Q^4) \\
\end{split}
\end{equation}
where $O( Q^4)$ stands for higher order mixed terms $Q_{b}^m Q_{f}^n,\, m+n \geq 4\, (m,n>0)$. Setting $t=q$, we recover (6).
\newline
\noindent\underline{$q$-Expansion} Up to overall factors of $t^{1/2}$ and/or $q^{1/2}$, we get
\begin{equation*}
\begin{split}
Q_b & : (1 + t + t^2) + (1 + 3 t + 3 t^2) q + (1 + 5 t + 7 t^2) q^2 + (1 + 7 t + 12 t^2) q^3 + (1 + 9 t + 19 t^2) q^4 \\
&  + (1 + 11 t + 27 t^2) q^5 + (1 + 13 t + 37 t^2) q^6 + (1 + 15 t + 48 t^2) q^7 + (1 + 17 t + 61 t^2) q^8 \\
& + (1 + 19 t + 75 t^2) q^9 + (1 + 21 t + 91 t^2) q^{10} + (1 + 23 t + 108 t^2) q^{11} + (1 + 25 t + 127 t^2) q^{12} \\
&  + (1 + 27 t + 147 t^2) q^{13} + (1 + 29 t + 169 t^2) q^{14} + (1 + 31 t + 192 t^2) q^{15} + O(q^{16}) \in \intg_{+}[t][[q]]\\
Q_{b}^2 & : (1 + t + t^2) q^2 + (1 + 3 t + 3 t^2) q^3 + (1 + 5 t + 7 t^2) q^4 + (1 + 7 t + 12 t^2) q^5 + (1 + 9 t + 19 t^2) q^6\\
& + (1 + 11 t + 27 t^2) q^7 + (1 + 13 t + 37 t^2) q^8 + (1 + 15 t + 48 t^2) q^9 + (1 + 17 t + 61 t^2) q^{10} \\
&  + (1 + 19 t + 75 t^2) q^{11} + (1 + 21 t + 91 t^2) q^{12} + (1 + 23 t + 108 t^2) q^{13} + (1 + 25 t + 127 t^2) q^{14}\\
& + (1 + 27 t + 147 t^2) q^{15} + O(q^{16}) \in \intg_{+}[t][[q]]\\
Q_b Q_f & : (t + t^2 + t^3) q + (1 + 2 t + 4 t^2 + 3 t^3) q^2 + (1 + 4 t + 8 t^2 + 7 t^3) q^3 + (1 + 6 t + 14 t^2 + 12 t^3) q^4\\
&  + (1 + 8 t + 21 t^2 + 19 t^3) q^5 + (1 + 10 t + 30 t^2 + 27 t^3) q^6 + (1 + 12 t + 40 t^2 + 37 t^3) q^7 \\
& + (1 + 14 t + 52 t^2 + 48 t^3) q^8+ (1 + 16 t + 65 t^2 + 61 t^3) q^9 + (1 + 18 t + 80 t^2 + 75 t^3) q^{10}\\
&  + (1 + 20 t + 96 t^2 + 91 t^3) q^{11} + (1 + 22 t + 114 t^2 + 108 t^3) q^{12} + (1 + 24 t + 133 t^2 + 127 t^3) q^{13}\\
&  + (1 + 26 t + 154 t^2 + 147 t^3) q^{14} + (1 + 28 t + 176 t^2 + 169 t^3) q^{15} + O(q^{16}) \in \intg_{+}[t][[q]]\\
\end{split}
\end{equation*}
Higher order $Q$-terms are recorded in the appendix.
\newline

\begin{conjecture} For a Hopf link $L$ colored by a $n$-dimensional skew-symmetric and/or $m$-dimensional symmetric representation $R$ of $su(2)$ in $\real P^3$, when specializing $Q_b$ and $Q_f$ to products of monomial factors in $q$ and $t$, a $q$-series expansion of $Z_{R \emptyset}(U,Q_b, Q_f, t,q)$ is a graded Poincare series of a colored $sl(N)$ link homology theory for links in $\real P^3$ up to an overall factor.  
\end{conjecture}

\section{Comparison with results of $S^3$}

In this section, we compare our results with that of $S^3$ or equivalently the resolved conifold in \cite{GIKV}. We note that the convention for the Young tableau used here is opposite of that of \cite{GIKV} (see Appendix A below).
\newline

In case of a colored link embedded in $S^3$, its invariant is a polynomial in $Q$ (i.e. the Khaler parameter) and its degree is set by the number of boxes of the Young tableau of the color. In $\real P^3$ case, we get a power series in $Q_b$ and $Q_f$. We can see from the above examples that the subsets of the terms corresponding to pure $Q_b$ terms are polynomials, whose degree is given by the number of boxes of the Young tableaux. For example, in case of () the leading and $Q_b$ terms are the same as that of the Hopf link in $S^3$ up to their relative sign (see Section 5.2 in \cite{GIKV}~\footnote{The change of variables are $t=q_1$ and $q=q_2$}). However, $Q_{b}^2$ term is different as there exists a combination of internal Young diagrams contributing, which does not exist for $S^3$ case at that order.
\newline

\indent A key difference between $S^3$ and $\real P^3$ is that  there are contributions by the mixed $Q_b Q_f$ terms even when the sum of their powers exceeds the total number of boxes in the Young tableaux, leading to the power series invariant. It is curious to find out physical and topological origins of $Q_b Q_f$ terms when their total powers are larger than the number of boxes $|\alpha| + |\gamma|$. In case of $S^3$ or equivalently the resolved conifold, the results of the unknot colored by the totally antisymmetric representations were independent of $t$ (see Section 5.1 and Appendix A in \cite{GIKV}) after the change of variable $Q(a)$. For our case, as a consequence of the presence of the mixed terms,  the results of the (totally) symmetric representations depends on $t$ (see Appendix C). This must hold for the potential change of variables for $Q_b$ and $Q_f$ in terms of $a_1$ and $a_2$. that is analogue of $Q(a)$.
\newline

\indent For colored unknots and Hopf links in $S^3$, it was shown that when the Kahler parameter $Q$ in $\hat{Z}_{\alpha\gamma^t}$ was specialized to $-\tilde{t} \tilde{q}^{-2N}$~\footnote{Further change of variables are $q_{1}^2=\tilde{t}^2 \tilde{q}^2, q_{2}=\tilde{q}^2$}, the rational function coefficients of $Q$-terms reduced to polynomials(see Appendix A.3 in \cite{GIKV}).   A speculation is that when $Q_b$ and $Q_f$ are specialized to products of monomials of $t$ and $q^{\pm N}$ in our $\hat{Z}_{\alpha\gamma^t}$, we expect that $\hat{Z}_{\alpha\gamma^t}$ would not reduce to (Laurent) polynomials in $t$ and $q$ due to the series nature of $\hat{Z}_{\alpha\gamma^t}$.\\

\noindent \textbf{Acknowledgments.} I would like to thank Pedro Guicardi, Sergei Gukov and Mrunmay Jagadale for discussions. I am grateful to Song Yu and Cumrun Vafa for reading a draft of this paper.   

\appendix
\section*{Appendix}
\addcontentsline{toc}{section}{Appendix}


\section{Conventions}

We summarize the 2d and 3d partitions and (skew) Schur functions and the conventions of the Young tableaux used in the paper; we follow the conventions used in \cite{IKV}.\\

\subsection{2d partitions}

A 2d partition is given by a Young tableaux $\nu= \lac \nu_1 \geq \nu_2 \geq \nu_3 \geq  \cdots | \nu_i \geq 0 \rac$, where $\nu_i$ is the number of boxes in $i$-th column. The size of $\nu$ is denoted by $|\nu|= \sum_{i} \nu_i $. The height of a Young tableaux decreases or stays the same. For example, $\nu= \lac 5,4,3,2,2,1 \rac $ corresponds to the first diagram above.
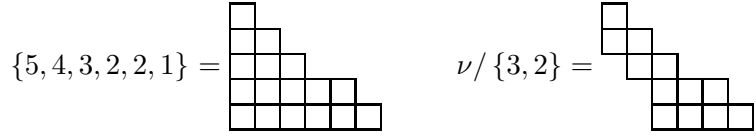
\begin{figure}[h]
$$ 
\lac 5,4,3,2,2,1 \rac = \ytableausetup{centertableaux} 
\begin{ytableau} 
~ \\
 ~ & ~ \\
 ~ & ~ & ~ \\
 ~ & ~ & ~ & ~ & ~ \\
 ~ & ~ & ~ & ~ & ~ & ~ \\  
\end{ytableau}
\hspace{1cm}
\nu / \lac 3,2 \rac = \ytableausetup{centertableaux} 
\begin{ytableau} 
~ \\
 ~ & ~ \\
\none & ~ & ~ \\
\none & \none & ~ & ~ & ~ \\
 \none & \none & ~ & ~ & ~ & ~ \\  
\end{ytableau}
$$
\end{figure}
\newline

\noindent From a Young tableaux its transpose can be defined as $\nu^t = \lac \nu_{1}^t , \nu_{2}^t , \nu_{3}^t ,  \cdots  \rac$. Let $(i,j) \in \nu$ be position of an upper right corner of a box, then $(j,i) \in \nu^t$. 
\newline

\noindent A subpartition $\lambda$ of $\nu$, if $(i,j) \in \lambda$  implies $(i,j) \in \nu$; it is denoted by $\lambda \subseteq \nu$. A skew partition denoted by $\nu /\lambda$ consists of all boxes of $\nu$, which are not in $\lambda$,
$$
\nu /\lambda = \lac (i,j) \in \nu | (i,j) \notin \lambda \rac.
$$
An example $\lac 5,4,3,2,2,1 \rac / \lac 3,2 \rac$ is shown above.

\subsection{3d partitions}

A 3d (plane) partition is a 3d generalization of the Young tableaux. The partition is an array of non-negative integers $\lac \pi_{i,j} | i,j \geq 1 \rac$ such that
$$
\pi_{i+r,j+s} \leq \pi_{i,j},\quad r,s \geq 0
$$
\noindent The partition consists of $\pi_{i,j}$ number of cubes at position $(i,j)$ stacked upwards. The total number of cubes is $|\pi|= \sum_{i,j} \pi_{i,j}$. An example of a 3d partition is given below.
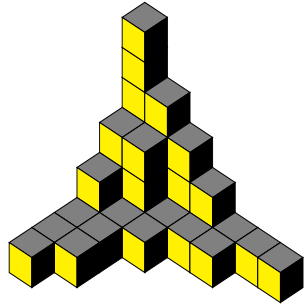
\begin{figure}[h]
\centering
\resizebox{4cm}{4cm}{
\begin{tikzpicture}
\planepartition{{6,4,3,2,1,1,1},{3,3,1,1,1},{2,1,1},{1,1},{1,1},{1}}
\end{tikzpicture}
}
\end{figure}
\newline
As in 2d partition, a skew partition can be defined in 3d. A 3d skew partition of shape $\nu/\lambda$ is an array of nonnegative integers $\lac \pi_{i,j} | (i,j) \in \nu/\lambda \rac$ such that
$$
\pi_{i+r,j+s} \leq \pi_{i,j},\quad r,s \geq 0.
$$
\noindent For a 3d skew partition whose boundary partitions given by $(\phi,\phi,\nu)$, there is a refined box counting function denoted by  $\tilde{Z}_{\nu} (t,q)$:
\begin{equation*}
\begin{split}
\tilde{Z}_{\nu} (t,q) & := \frac{Z_{\nu} (t,q)}{Z_{\phi} (t,q)}\\
\tilde{Z}_{\nu} (t,q) & =  \prod_{s \in \nu } \lb 1- t^{a(s)+1} q^{l(s)} \rb^{-1} = \prod_{s \in \nu^t } \lb 1- t^{l(s)+1} q^{a(s)} \rb^{-1}\\
\tilde{Z}_{\nu} (t,q) & = t^{-\frac{||\nu ||^2}{2}} P_{\nu^t}(t^{-\rho};q,t)\\
\end{split}
\end{equation*}
where $a(s)=a(i,j)$ is number of boxes on the right of the $(i,j)$-box and $l(s)=l(i,j)$ is number of boxes on the top of the $(i,j)$-box. They are related to the hook length of $(i,j)$-box as $a(s)+l(s)+1$. $P_{\nu^t}(\textbf{z};q,t)$ is the symmetric Macdonald function and $\tilde{Z}_{\nu} (t,q)$ counts number of boxes in the Young diagram $\nu$. Setting $t=q$ reduces to the regular $\tilde{Z}_{\nu} (q)$.

\section{Expansions}

We record the normalization $Z_{\phi\phi} (Q_b, Q_f ,t,q)$. We note that a different final form of it was computed in \cite{IKV}. That final form is not suitable for our purpose.

\begin{equation*}
\begin{split}
Z_{\phi\phi} & = 1 + \frac{1}{(1-t)(1-q)} \,\lsb 2 Q_b \, \sqrt{tq} + Q_f \, (t+q) \rsb + \frac{1}{\sqrt{tq}(1-q)(1-q^2)(1-t)(1-t^2)} \,\lsb Q_{b}^2 \,(tq)^{3/2}\right.\\
& \left. \times (3 + q + t + 3 q t) + Q_b Q_f \, (t^2 - t^4 + 2 q t (1 + t) + 2 q^3 t^2 (1 + t) + q^4 (-1 + t^2) + q^2 (1 + t)^2 (1 + t^2))\right.\\
& \left. + Q_{f}^2 \,\sqrt{tq} (q^3 t + t^2 + q^2 (1 + t + t^2) + q t (1 + t + t^2)) \rsb\\
&  + \frac{1}{(tq)^{3/2}(1-q)(1-q^2)(1-q^3)(1-t)(1-t^2)(1-t^3)} \, \lsb \, Q_{b}^3 \,\lb 2 q^4 t^3 (1 + t)^3 + 2 q^5 t^3 (1 + t)^3 \right.\right.\\
& \left.\left. + 2 q^3 t^3 (2 + t + t^2) + 2 q^6 t^4 (1 + t + 2 t^2) \rb + Q_{b}^2 Q_f \,\sqrt{tq} (1 + q + q^2) (1 + t + t^2) \lb q^6 (-1 + t)^2 (1 + t)\right.\right.\\
& \left.\left. + (-1 + t)^2 t^3 (1 + t) - q t^2 (-2 + t + t^2 - t^3 + t^4) + q^5 (-1 + t - t^2 - t^3 + 2 t^4)\right.\right.\\
& \left.\left. - q^2 t (-2 - 2 t^2 - 2 t^3 + t^4 + t^5) + q^4 (-1 - t + 2 t^2 + 2 t^3 + 2 t^5) + q^3 (1 - t + 2 t^2 + 4 t^3 + 2 t^4 - t^5 + t^6)\rb\right.\\
& \left. + Q_b Q_f^{2} \, (1 + q + q^2) (1 + t + t^2) \lb q^7 (-1 + t)^2 (1 + t) + (-1 + t)^2 t^4 (1 + t) + q^6 (-1 + t - t^3 + t^4) \right.\right.\\
& \left. \left. - q t^3 (-1 + t - t^3 + t^4) + q^5 (-1 + t^2 + 2 t^5) + q^4 (1 - t + 3 t^3 + 2 t^4 + t^6) + q^2 (2 t^2 + t^5 - t^7) \right.\right.\\
& \left. \left. + q^3 (t + 2 t^3 + 3 t^4 - t^6 + t^7) \rb + Q_f^{3} \,\lb q^{3/2} t^{9/2} + q^{15/2} t^{9/2} + q^{13/2} t^{5/2} (1 + 2 t + t^2 + t^3)\right. \right.\\
& \left.\left. + q^{5/2} t^{7/2} (1 + t + 2 t^2 + t^3) + q^{11/2} t^{5/2} (2 + 3 t + 3 t^2 + 2 t^3 + t^4) \right.\right.\\
& \left. \left. + q^{7/2} t^{5/2} (1 + 2 t + 3 t^2 + 3 t^3 + 2 t^4) + q^{9/2} t^{3/2} (1 + t + 3 t^2 + 4 t^3 + 3 t^4 + t^5 + t^6)\rb \rsb + O(Q^4).
\end{split}
\end{equation*}
This expression is symmetric under $t \leftrightarrow q$. For $t=q$, it reduces to the regular normalization factor $Z_{\phi\phi} (Q_b, Q_f ,q)$.
\newline

\noindent We list the $q$ expansions of the colored unknots and Hopf links in Section 4.
\newline

\noindent $Z_{\Lambda^2 \emptyset}(\vec{Q},t,q)$
\begin{equation*}
\begin{split}
Q_b Q_f^2 & : t^2 + t^3 +  (t + 2 t^2 + 2 t^3)q +  (1 + 2 t + 3 t^2 + 3 t^3)q^2 + (1 + 3 t + 4 t^2 + 4 t^3) q^3\\
&  + (1 + 4 t + 5 t^2 + 5 t^3)q^4  + (1 + 5 t + 6 t^2 + 6 t^3) q^5  + (1 + 6 t + 7 t^2 + 7 t^3)q^6 \\
& +  (1 + 7 t + 8 t^2 + 8 t^3)q^7  +  (1 + 8 t + 9 t^2 + 9 t^3)q^8 + (1 + 9 t + 10 t^2 + 10 t^3) q^9 \\
&  +  (1 + 10 t + 11 t^2 + 11 t^3)q^{10} +  (1 + 11 t + 12 t^2 + 12 t^3)q^{11} +  (1 + 12 t + 13 t^2 + 13 t^3) q^{12}\\
&  +   (1 + 13 t + 14 t^2 + 14 t^3)q^{13} +  (1 + 14 t + 15 t^2 + 15 t^3)q^{14}  +  (1 + 15 t + 16 t^2 + 16 t^3)q^{15}\\
&  + O(q^{16}) \in \intg_{+}[t][[q]]\\
Q_b^2 Q_f & : t^2 + t^3 + 2 (t + 2 t^2 + t^3)q  +  (1 + 4 t + 7 t^2 + 3 t^3)q^2 +   (1 + 6 t + 10 t^2 + 4 t^3)q^3\\
& + (1 + 8 t + 13 t^2 + 5 t^3) q^4 +   (1 + 10 t + 16 t^2 + 6 t^3)q^5 +  (1 + 12 t + 19 t^2 + 7 t^3)q^6 \\
& +   (1 + 14 t + 22 t^2 + 8 t^3)q^7 + (1 + 16 t + 25 t^2 + 9 t^3)  q^8 +  (1 + 18 t + 28 t^2 + 10 t^3)q^9  \\
& +  (1 + 20 t + 31 t^2 + 11 t^3) q^{10} +   (1 + 22 t + 34 t^2 + 12 t^3) q^{11} +   (1 + 24 t + 37 t^2 + 13 t^3)q^{12} \\
& +  (1 + 26 t + 40 t^2 + 14 t^3)q^{13}  +  (1 + 28 t + 43 t^2 + 15 t^3)  q^{14} + (1 + 30 t + 46 t^2 + 16 t^3)q^{15}\\
&  + O(q^{16}) \in \intg_{+}[t][[q]]\\
\end{split}
\end{equation*}
\newline

\noindent $Z_{\Lambda^3 \emptyset}(\vec{Q},t,q)$
\begin{equation*}
\begin{split}
Q_b^2 Q_f & : t^2 (1 + t + t^2) + t (3 + 5 t + 5 t^2 + 2 t^3)q  + (2 + 6 t + 11 t^2 + 10 t^3 + 4 t^4)q^2  \\
& +(2 + 9 t + 18 t^2 + 18 t^3 + 6 t^4) q^3  + (2 + 12 t + 27 t^2 + 27 t^3 + 9 t^4)q^4  \\
& + (2 + 15 t + 37 t^2 + 39 t^3 + 12 t^4)q^5  +  (2 + 18 t + 49 t^2 + 52 t^3 + 16 t^4)q^6  \\
& + (2 + 21 t + 62 t^2 + 68 t^3 + 20 t^4) q^7  +  (2 + 24 t + 77 t^2 + 85 t^3 + 25 t^4) q^8\\
& + (2 + 27 t + 93 t^2 + 105 t^3 + 30 t^4)q^9  + (2 + 30 t + 111 t^2 + 126 t^3 + 36 t^4)q^{10}  + O(q^{11}) \in \intg_{+}[t][[q]]\\
Q_b Q_f^2 & : t^2 + t^3 + t^4 + (t + 2 t^2 + 3 t^3 + 2 t^4)q  + (1 + 2 t + 4 t^2 + 5 t^3 + 4 t^4) q^2\\
& +(1 + 3 t + 6 t^2 + 8 t^3 + 6 t^4) q^3  + (1 + 4 t + 9 t^2 + 11 t^3 + 9 t^4)q^4  +(1 + 5 t + 12 t^2 + 15 t^3 + 12 t^4) q^5 \\
& +(1 + 6 t + 16 t^2 + 19 t^3 + 16 t^4) q^6  +  (1 + 7 t + 20 t^2 + 24 t^3 + 20 t^4)q^7 \\
&  + (1 + 8 t + 25 t^2 + 29 t^3 + 25 t^4)q^8  +(1 + 9 t + 30 t^2 + 35 t^3 + 30 t^4) q^9 \\
& + (1 + 10 t + 36 t^2 + 41 t^3 + 36 t^4)  q^{10} + O(q^{11}) \in \intg_{+}[t][[q]]\\
\end{split}
\end{equation*}
\newline

\noindent $Z_{\square \square}(\vec{Q},t,q)$
\begin{equation*}
\begin{split}
Q_b^2 Q_f & : t^2 (1 + t) +  t (2 + 5 t + 3 t^2)q +  (1 + 5 t + 11 t^2 + 5 t^3)q^2 +   (1 + 9 t + 17 t^2 + 7 t^3)q^3 \\
& + (1 + 13 t + 23 t^2 + 9 t^3)q^4  +  (1 + 17 t + 29 t^2 + 11 t^3)q^5  + (1 + 21 t + 35 t^2 + 13 t^3)q^6  \\
& +   (1 + 25 t + 41 t^2 + 15 t^3)q^7  +   (1 + 29 t + 47 t^2 + 17 t^3)q^8 +  (1 + 33 t + 53 t^2 + 19 t^3)q^9 \\
& +   (1 + 37 t + 59 t^2 + 21 t^3) q^{10} +  (1 + 41 t + 65 t^2 + 23 t^3) q^{11} +   (1 + 45 t + 71 t^2 + 25 t^3)q^{12}\\
& +   (1 + 49 t + 77 t^2 + 27 t^3)q^{13} +  (1 + 53 t + 83 t^2 + 29 t^3)q^{14}  + (1 + 57 t + 89 t^2 + 31 t^3)q^{15} \\
& + O(q^{16}) \in \intg_{+}[t][[q]]\\
Q_b Q_f^2 & : t^2 + t^3 +  (t + 2 t^2 + 3 t^3)q + (1 + 2 t + 4 t^2 + 5 t^3)q^2  +  (1 + 4 t + 6 t^2 + 7 t^3)q^3\\
& +(1 + 6 t + 8 t^2 + 9 t^3) q^4  + (1 + 8 t + 10 t^2 + 11 t^3) q^5  +(1 + 10 t + 12 t^2 + 13 t^3) q^6 \\
& +  (1 + 12 t + 14 t^2 + 15 t^3)q^7  + (1 + 14 t + 16 t^2 + 17 t^3) q^8  +  (1 + 16 t + 18 t^2 + 19 t^3)q^9 \\
& +   (1 + 18 t + 20 t^2 + 21 t^3)q^{10} +  (1 + 20 t + 22 t^2 + 23 t^3)  q^{11} +   (1 + 22 t + 24 t^2 + 25 t^3)q^{12}\\
& +   (1 + 24 t + 26 t^2 + 27 t^3)q^{13} +  (1 + 26 t + 28 t^2 + 29 t^3) q^{14} +  (1 + 28 t + 30 t^2 + 31 t^3)q^{15}\\
& + O(q^{16}) \in \intg_{+}[t][[q]]\\
\end{split}
\end{equation*}
\newline

\noindent  $Z_{\square \Lambda^2 }(\vec{Q},t,q)$
\begin{equation*}
\begin{split}
Q_{b}^3 & : 1 + 2 q + 4 q^2 + 6 q^3 + 9 q^4 + 12 q^5 + 16 q^6 + 20 q^7 + 25 q^8 + 30 q^9 + 36 q^{10} + 42 q^{11} + 49 q^{12}\\
& + 56 q^{13} + 64 q^{14} + 72 q^{15}  + O(q^{16}) \in \intg_{+}[t][[q]]\\
Q_b^2 Q_f & : t^2 (1 + t + t^2) +  3 t (1 + 2 t + 2 t^2 + t^3) q + (2 + 7 t + 16 t^2 + 16 t^3 + 7 t^4) q^2 \\
& + (2 + 13 t + 30 t^2 + 33 t^3 + 12 t^4) q^3 + (2 +  19 t + 49 t^2 + 55 t^3 + 19 t^4) q^4\\
&  + (2 + 25 t + 72 t^2 + 84 t^3 + 27 t^4) q^5 + (2 + 31 t + 100 t^2 + 118 t^3 + 37 t^4) q^6 \\
& + (2 + 37 t + 132 t^2 + 159 t^3 + 48 t^4) q^7+ (2 + 43 t + 169 t^2 + 205 t^3 + 61 t^4) q^8\\
& + (2 + 49 t + 210 t^2 + 258 t^3 + 75 t^4) q^9 + (2 + 55 t + 256 t^2 + 316 t^3 + 91 t^4) q^{10}\\
& + (2 + 61 t + 306 t^2 + 381 t^3 + 108 t^4) q^{11} + (2 + 67 t + 361 t^2 + 451 t^3 +  127 t^4) q^{12}\\
& + (2 + 73 t + 420 t^2 + 528 t^3 + 147 t^4) q^{13}  + (2 + 79 t + 484 t^2 + 610 t^3 + 169 t^4) q^{14}\\
& + (2 + 85 t + 552 t^2 + 699 t^3 + 192 t^4) q^{15} + O(q^{16}) \in \intg_{+}[t][[q]]\\
Q_b Q_f^2 & : (t^2 + t^3 +  t^4) + (t + 2 t^2 + 4 t^3 + 3 t^4) q + (1 + 2 t + 5 t^2 + 8 t^3 + 7 t^4) q^2\\
& + (1 + 4 t + 9 t^2 + 14 t^3 + 12 t^4) q^3 + (1 + 6 t + 15 t^2 + 21 t^3 + 19 t^4) q^4\\
& + (1 + 8 t + 22 t^2 + 30 t^3 + 27 t^4) q^5 + (1 + 10 t + 31 t^2 + 40 t^3 + 37 t^4) q^6\\
&  + (1 + 12 t + 41 t^2 + 52 t^3 + 48 t^4) q^7 + (1 + 14 t + 53 t^2 + 65 t^3 + 61 t^4) q^8 \\
&  + (1 + 16 t + 66 t^2 + 80 t^3 + 75 t^4) q^9 + (1 + 18 t + 81 t^2 + 96 t^3 + 91 t^4) q^{10}\\
&  + (1 + 20 t + 97 t^2 + 114 t^3 + 108 t^4) q^{11} + (1 + 22 t + 115 t^2 + 133 t^3 + 127 t^4) q^{12}\\
&  + (1 + 24 t + 134 t^2 + 154 t^3 + 147 t^4) q^{13} + (1 + 26 t + 155 t^2 + 176 t^3 + 169 t^4) q^{14}\\
&  + (1 + 28 t + 177 t^2 + 200 t^3 + 192 t^4) q^{15} + O(q^{16}) \in \intg_{+}[t][[q]]\\
\end{split}
\end{equation*}

\section{Other Representations}

\noindent For 2d symmetric representation $S^2 = \ytableausetup{smalltableaux}\begin{ytableau}  ~ &   \end{ytableau} $.
\begin{equation*}
\begin{split}
Z_{S^2 \emptyset}(\vec{Q},t,q) & = \frac{q^2}{(1-q) \left(1-q^2\right)} + Q_b \, \frac{q^{5/2}}{t^{1/2}(1-q)^2} + \frac{q^{3/2}}{(1-q) \left(1-q^2\right)}\lsb Q_{b}^2 \, q^{3/2} + Q_b Q_f \,t^{1/2}(1+q)(q+t) \rsb \\
& + \frac{q^{1/2}}{t(1-q)^2} \lsb    Q_{b}^2 Q_f \,(q^{3/2}+q^{5/2}+2 q^{3/2} t+\sqrt{q} t (1+t)) +  Q_b Q_{f}^2 \,(q^2 \sqrt{t}+q t^{3/2}+t^{5/2})  \rsb +  O( Q^4)\\
\end{split}
\end{equation*}
where $O( Q^4)$ stands for higher order mixed terms $Q_{b}^m Q_{f}^n,\, m+n \geq 4\, (m,n>0)$.

\begin{equation*}
\begin{split}
Q_b Q_f & : t + (1 + 2 t) q + (2 + 3 t) q^2 + (3 + 4 t) q^3 + (4 + 5 t) q^4 + (5 + 6 t) q^5 + (6 + 7 t) q^6 \\
& + (7 + 8 t) q^7 + (8 + 9 t) q^8 + (9 + 10 t) q^9 + (10 + 11 t) q^{10} + (11 +   12 t) q^{11} + (12 + 13 t) q^{12} \\
& + (13 + 14 t) q^{13} + (14 +     15 t) q^{14} + (15 + 16 t) q^{15} + O(q^{16}) \in \intg_{+}[t][[q]]\\
Q_b Q_f^2 & : t^2 + (t + 2 t^2) q + (1 + 2 t + 3 t^2) q^2 + (2 + 3 t + 4 t^2) q^3 + (3 + 4 t + 5 t^2) q^4\\
& + (4 + 5 t + 6 t^2) q^5 + (5 + 6 t + 7 t^2) q^6 + (6 + 7 t + 8 t^2) q^7 + (7 + 8 t + 9 t^2) q^8\\
& + (8 + 9 t + 10 t^2) q^9 + (9 + 10 t + 11 t^2) q^{10} + (10 + 11 t + 12 t^2) q^{11} + (11 + 12 t +  13 t^2) q^{12}\\
& + (12 + 13 t + 14 t^2) q^{13} + (13 + 14 t +  15 t^2) q^{14} + (14 + 15 t + 16 t^2) q^{15}  + O(q^{16}) \in \intg_{+}[t][[q]]\\
Q_b^2 Q_f & : (t + t^2) + (1 + 4 t + 2 t^2) q + (3 + 7 t + 3 t^2) q^2 + (5 + 10 t + 4 t^2) q^3\\
& + (7 + 13 t + 5 t^2) q^4 + (9 + 16 t + 6 t^2) q^5 + (11 + 19 t + 7 t^2) q^6 + (13 + 22 t +  8 t^2) q^7\\
& + (15 + 25 t + 9 t^2) q^8 + (17 + 28 t +  10 t^2) q^9 + (19 + 31 t + 11 t^2) q^{10} \\
& + (21 + 34 t +  12 t^2) q^{11} + (23 + 37 t + 13 t^2) q^{12} + (25 + 40 t +  14 t^2) q^{13} \\
& + (27 + 43 t + 15 t^2) q^{14} + (29 + 46 t + 16 t^2) q^{15}  + O(q^{16}) \in \intg_{+}[t][[q]]\\
\end{split}
\end{equation*}

\noindent For 3d symmetric representation $S^3 = \ytableausetup{smalltableaux}\begin{ytableau}  ~ & ~ &  \end{ytableau} $.

\begin{equation*}
\begin{split}
Z_{S^3 \emptyset}(\vec{Q},t,q) & = \frac{q^{9/2}}{(1-q) (1-q^2)(1-q^3)} +  Q_b \, \frac{q^5}{t^{1/2}(1-q)^2 (1-q^2)} + \frac{1}{t(1-q)^2 (1-q^2)}\lsb Q_{b}^2 q^{11/2} \right.\\
& \left. +  Q_b Q_f (q^5 t^{1/2} + q^4 t^{3/2} ) \rsb + \frac{q^3}{t^{3/2} (1-q) (1-q^2)(1-q^3)} \lsb Q_{b}^3 \, q^3  +  Q_{b}^2 Q_f \,  \left(2 q^{9/2} \sqrt{t} \right.\right.\\
& \left.\left. +3 q^{7/2} \sqrt{t} (1+t)+\sqrt{q} t^{3/2} (1+t)+q^{3/2} \sqrt{t} \left(1+4 t+t^2\right)+q^{5/2} \sqrt{t} \left(3+4 t+t^2\right)\right) \right.\\
& \left. +  Q_b Q_{f}^2 \,\left(q^4 t+t^3 +q^3 t (1+t) +q t^2 (1+t)+q^2 t \left(1+t+t^2\right)\right) \rsb +  O( Q^4)\\
\end{split}
\end{equation*}
where $O( Q^4)$ stands for higher order mixed terms $Q_{b}^m Q_{f}^n,\, m+n \geq 4\, (m,n>0)$. We observe that these representations also yield positive integer coefficients in their $q$ expansions.

\begin{equation*}
\begin{split}
Q_b Q_f & : t+(1+2 t) q+(2+4 t) q^2+(4+6 t) q^3+(6+9 t) q^4+(9+12 t) q^5+(12+16 t) q^6+(16+20 t) q^7\\
& +(20+25 t) q^8+(25+30 t)q^9+(30+36 t) q^{10}+(36+42 t) q^{11}+(42+49 t) q^{12}+(49+56 t) q^{13}\\
& +(56+64 t) q^{14}+(64+72 t)q^{15} + O(q^{16}) \in \intg_{+}[t][[q]]\\
Q_b Q_f^2 & : t^2+t (1+2 t) q+\left(1+2 t+4 t^2\right) q^2+\left(2+4 t+6 t^2\right) q^3+\left(4+6 t+9 t^2\right) q^4 +\left(6+9 t+12 t^2\right) q^5\\
&+\left(9+12 t+16 t^2\right) q^6 +4 \left(3+4 t+5 t^2\right) q^7 +\left(16+20 t+25 t^2\right) q^8 +5\left(4+5 t+6 t^2\right) q^9\\
& +\left(25+30 t+36 t^2\right) q^{10}+6 \left(5+6 t+7 t^2\right) q^{11}+\left(36+42 t+49 t^2\right) q^{12} +7 \left(6+7 t+8 t^2\right) q^{13}\\
& +\left(49+56 t+64 t^2\right) q^{14}+8 \left(7+8 t+9 t^2\right) q^{15} + O(q^{16}) \in \intg_{+}[t][[q]]\\
Q_b^2 Q_f & :t (1+t)+\left(1+5 t+2 t^2\right) q+\left(4+10 t+4 t^2\right) q^2+\left(8+18 t+6 t^2\right) q^3 +\left(14+27 t+9t^2\right) q^4\\
& +3 \left(7+13 t+4 t^2\right) q^5+\left(30+52 t+16 t^2\right) q^6 +\left(40+68 t+20 t^2\right)q^7+\left(52+85 t+25 t^2\right) q^8\\
&+5 \left(13+21 t+6 t^2\right) q^9+\left(80+126 t+36 t^2\right) q^{10}+6\left(16+25 t+7 t^2\right) q^{11} +\left(114+175 t+49 t^2\right) q^{12}\\
& +7 \left(19+29 t+8 t^2\right) q^{13}+\left(154+232 t+64 t^2\right) q^{14}+8 \left(22+33 t+9 t^2\right) q^{15}  + O(q^{16}) \in \intg_{+}[t][[q]]\\
\end{split}
\end{equation*}


	%
	%
	%
	%
	%

\end{document}